\documentclass[prl,twocolumn,nofootinbib,superscriptaddress,showpacs,aps,10pt]{revtex4-1}

\usepackage{amsmath}
\usepackage{amssymb}
\usepackage{graphicx}
\usepackage{dcolumn,dsfont}
\usepackage{bm,braket}
\usepackage{hyperref}
\usepackage{isotope}
\usepackage{multirow}
\usepackage{tabularx}
\usepackage{comment}
\usepackage{url}
\newcommand{\MeV}{\,\mathrm{MeV}}

\newcommand{\fmiq}{\,\mathrm{fm}^{-3}}
\newcommand{\gcmiq}{\,\mathrm{g~cm}^{-3}}
\newcommand{\Gth}{\Gamma_\mathrm{th}}

\begin{document}

\title{Equation of state effects in core-collapse supernovae}

\author{H. Yasin}
\email[Email:~]{hannah.yasin@physik.tu-darmstadt.de}
\affiliation{Institut f\"ur Kernphysik, Technische Universit\"at Darmstadt, 64289 Darmstadt, Germany}
\author{S. Sch\"afer}
\email[Email:~]{sschaefer@theorie.ikp.physik.tu-darmstadt.de}
\affiliation{Institut f\"ur Kernphysik, Technische Universit\"at Darmstadt, 64289 Darmstadt, Germany}
\affiliation{ExtreMe Matter Institute EMMI, GSI Helmholtzzentrum f\"ur Schwerionenforschung GmbH, 64291 Darmstadt, Germany}
\author{A. Arcones}
\email[Email:~]{almudena.arcones@physik.tu-darmstadt.de}
\affiliation{Institut f\"ur Kernphysik, Technische Universit\"at Darmstadt, 64289 Darmstadt, Germany}
\affiliation{GSI Helmholtzzentrum f\"ur Schwerionenforschung GmbH, 64291 Darmstadt, Germany}
\author{A. Schwenk}
\email[Email:~]{schwenk@physik.tu-darmstadt.de}
\affiliation{Institut f\"ur Kernphysik, Technische Universit\"at Darmstadt, 64289 Darmstadt, Germany}
\affiliation{ExtreMe Matter Institute EMMI, GSI Helmholtzzentrum f\"ur Schwerionenforschung GmbH, 64291 Darmstadt, Germany}
\affiliation{Max-Planck-Institut f\"ur Kernphysik, Saupfercheckweg 1, 69117 Heidelberg, Germany}

\begin{abstract}
We investigate the impact of different properties of the nuclear
equation of state in core-collapse supernovae, with a focus on the
proto-neutron-star contraction and its impact on the shock
evolution. To this end, we introduce a range of equations of state
that vary the nucleon effective mass, incompressibility, symmetry
energy, and nuclear saturation point. This allows us to point to the
different effects in changing these properties from the Lattimer and
Swesty to the Shen {\it et al.} equations of state, the two most
commonly used equations of state in simulations.  In particular, we
trace the contraction behavior to the effective mass, which determines
the thermal nucleonic contributions to the equation of state. Larger
effective masses lead to lower pressures at nuclear densities and a
lower thermal index. This results in a more rapid contraction of the
proto-neutron star and consequently higher neutrino energies, which
aids the shock evolution to a faster explosion.
\end{abstract}

\maketitle

Core-collapse supernovae and neutron star mergers are cosmic
laboratories for physics at the extremes.
In the new multimessenger era, including also
gravitational wave detection~\cite{Abbott:2018exr}, we can uniquely
combine observations and hydrodynamic simulations to learn more about
these events. One critical microphysics input in simulations is the 
equation of state (EOS). In this Letter, we explore the macroscopic
effects of the microphysics in the EOS in the context of supernova explosions.

Massive stars end their lives as core-collapse supernovae when their
central iron cores collapse forming a proto-neutron star (PNS) and a
shock wave that propagates through the infalling stellar layers. The
final success of the shock to destroy the star depends on the neutrino
energy deposited behind the shock. This is affected by convection,
hydrodynamic instabilities, rotation, magnetic fields, and by the
evolution of the PNS.  Despite the many advances in simulating
core-collapse supernovae including three-dimensional simulations (see, e.g.,
Refs.~\cite{Kotake2012,Burrows2013,Lentz2015, Mueller2016,Janka:2016fox,Bollig2017,Evan2018Comparison}), the details about the explosion are still not clear.

The EOS is constrained by modern theoretical calculations at nuclear
densities~\cite{Hebe10nmatt,Gand12nm,Hebe13ApJ,Krue13N3LOlong,Holt13PPNP,Carb13nm,Hage14ccnm,Well14nmtherm,Dris16asym,Lynn16QMC3N,Tews18cs,Drischler:2017wtt},
by nuclear experiments (see, e.g.,
Ref.~\cite{Tsan12esymm,LattimerLim,Birk17dipole}) as well as
observations, in particular of two-solar-mass neutron
stars~\cite{Anto13PSRM201,Fonseca2016}. However, the properties of
the EOS at densities above $(1-2) n_0$ (with saturation density $n_0
\approx 0.16 \fmiq$) remain quite uncertain, but these are relevant for simulations.  There are two ``classical'' and commonly used EOSs in
tabulated form, which cover the broad range of conditions reached in
supernova simulations: the Lattimer and Swesty (LS)
EOS~\cite{Latt91LSEOS,Latt85LLPR} and the H.~Shen \textit{et al.}~(Shen)
EOS~\cite{Shen98EOS}. Recently, there have been major efforts to
provide new EOS tables (see, e.g.,
Refs.~\cite{Hempel2010,GShen2011,Hempel.etal:2012,Steiner:2012rk,Schneider2017}).

A major impact of the EOS in supernova simulations is due to
variations in the PNS contraction (see, e.g.,
Refs.~\cite{MarekJanka2009,Hempel.etal:2012}).  A faster contraction
during the first few hundred milliseconds after bounce favors
explosions due to higher neutrino energies and thus increased
heating~\cite{MarekJanka2009,Janka2012review}. This has been discussed
when comparing different EOSs (including LS and
Shen)~\cite{Sumi2005,MarekJanka2009,Janka2012review,Couch2013,Suwa2013}.
However, these studies are usually performed based on EOSs that differ
in their underlying theoretical framework (based on
Skyrme at high densities~\cite{Latt91LSEOS,Schneider2017} or relativistic energy-density
functionals \cite{Shen98EOS,Hempel2010,GShen2011, Hempel2010,Hempel.etal:2012,Steiner:2012rk})
or within the same framework, varying all EOS parameters
simultaneously~\cite{Hempel2010,Hempel.etal:2012,Steiner:2012rk,Schneider2017}.
This makes it difficult to link the behavior of the PNS and shock to a
particular nuclear physics input. The only EOS work where solely one parameter 
was changed are those based on LS with different
incompressibilities~\cite{Latt91LSEOS}, which were applied, 
e.g., in Refs.~\cite{Suwa2013, Couch2013}. In this Letter, we
individually vary different nuclear matter properties within the same
EOS framework to clearly identify the impact of the effective mass,
incompressibility, symmetry energy, and saturation point on the
physics of core-collapse supernovae.

{\it Equation of state and supernova simulations.--} The LS EOS is based 
on a Skyrme energy-density functional, where the energy per nucleon
of uniform matter as a function of baryon density $n$ and proton fraction 
$x = n_p/n$ at zero temperature is given by~\cite{Latt91LSEOS}
\begin{align}
\frac{E}{A}\biggr|_{T=0} &= \frac{3\hbar^2}{10m^*} (3\pi^2n)^{2/3}
\Bigl[(1-x)^{5/3}+x^{5/3}\Bigr] \nonumber \\ 
&\quad + \Bigl[a+4bx(1-x)\Bigr] n + cn^\delta - x\Delta \,.
\end{align}
Here, $a, b, c$, and $\delta$ are the Skyrme parameters, and $\Delta$
is the neutron-proton mass difference. The nucleon effective mass $m^*$ is given by $\hbar^2/(2m^*) = \hbar^2/(2m) + \alpha n$, with $m=m_n =m_p=939.5654 \MeV$ in LS, and $\alpha$ is fit to the effective mass at saturation density.

\begin{table}[t]
\begin{center}
\begin{tabular}{l|cccc|cc}
\hline \hline
& $m^*/m$ & $K$ & $E_\mathrm{sym}$ & $L$ & $n_0$ & $B$ \\ \hline
LS220 & 1.0 & 220 & 29.6 & 73.7 & 0.155 & 16.0 \\
Shen & 0.634 & 281 & 36.9\footnote{The symmetry energy in Shen is obtained via the second derivative of the energy per particle and not from the difference of neutron and symmetric matter as in LS.} & 110.8 & 0.145 & 16.3 \\
Theo. & $0.9(2)$ & $215(40)$ & $32(4)$ & $51(19)$ & $0.164(7)$ & $15.86(57)$ \\
\hline\hline
\end{tabular}
\caption{Parameters for the LS220 and Shen EOS compared to
theoretical ranges (``Theo.") from chiral EFT calculations for the
effective mass $m^*$ at saturation density~\cite{Hebe09enerfunc,Well14nmtherm,Dris16gap},
incompressibility $K$~\cite{Hebe11fits,Dris16asym},
symmetry energy $E_\mathrm{sym}$~\cite{Hebe13ApJ,Dris16asym}, 
and $L$ parameter~\cite{Hebe13ApJ,Drischler:2017wtt} as well as
the empirical ranges for the saturation density $n_0$ and energy $B$ 
given by the compilation in Ref.~\cite{Drischler:2017wtt}. All dimensionful 
quantities are in MeV except $n_0$ is in $\fmiq$.\label{tab:parameters}}
\end{center}
\end{table}

In Table~\ref{tab:parameters}, we list the EOS parameters for the LS EOS
with incompressibility $K=220 \MeV$ (LS220) and the Shen EOS. We
choose LS220 from the LS family, as this EOS supports a two-solar-mass
neutron star and the incompressibility lies within the expected range
from nuclear physics (see Table~\ref{tab:parameters}).  Moreover, 
Table~\ref{tab:parameters} includes theoretical ranges from chiral EFT
calculations and from the extraction of the empirical saturation
point. As the effective mass is expected to be reduced at saturation
density, we explore the impact of the effective mass by changing this
from $m^*=m$ (LS220) to $m^*/m=0.8$ to $m^*/m=0.634$ (the Shen
value). For the latter scenarios, we refit the Skyrme parameters $a,
b, c$, and $\delta$ for given $m^*$ to reproduce the same saturation
density $n_0$ and energy $B$, the incompressibility $K$, and symmetry
energy $E_\mathrm{sym}$.  This defines EOSs that
are labeled as $m^*_{0.8}$ and $m^*_\mathrm{S}$, respectively. On top
of $m^*_\mathrm{S}$, we vary the incompressibility
$(m^*,K)_\mathrm{S}$, symmetry energy
$(m^*,E_\mathrm{sym})_\mathrm{S}$, and both
$(m^*,K,E_\mathrm{sym})_\mathrm{S}$ to the values of the Shen EOS.
The EOS labeled SkShen additionally uses Shen values for saturation
density and energy. In each case, the Skyrme parameters are refit so
that the EOS parameters are varied one at a time. Finally, we note that
the $L$ parameter, which determines the pressure of neutron matter, is
not an independent parameter in the LS Skyrme functional (because
there is only an isoscalar density-dependent $c$ term), but is
determined by the other nuclear matter properties, such that the $L$
parameter varies for all constructed EOSs between the values of LS220 and Shen.

The EOS tables are created using the open-source code SROEOS from
Ref.~\cite{Schneider2017,sroeos}. As a check, we also implemented an
effective mass in the original code from
LS~\cite{Latt91LSEOS,lseosonline}. Both codes agree very well, except
for small differences within the phase transition region (also
discussed in Ref.~\cite{Schneider2017}), but we have checked that
these do not affect the findings of this Letter. As in Ref.~\cite{Schneider2017},
we refer to the LS220 generated EOS with the SROEOS code as 
LS220$^\dagger$. The Shen EOS table is taken
from Ref.~\cite{OConnorOtt2010,oconnertable}.

The constructed EOSs offer novel insights into the impact of individual
nuclear physics input on core-collapse supernovae. We perform
spherically symmetric simulations with the FLASH
code~\cite{Fryxell2000} for a 15~$M_\odot$
progenitor~\cite{WHW02}. Moreover, we use the two-moment,
energy-dependent, multispecies, neutrino radiation transport 
scheme M1 with an analytic
closure~\cite{Shibata2011,OConnor2015,OConnorCouch2018},
applying the standard neutrino rates of Ref.~\cite{NuLibOnline}, 
which include 
neutrino scattering on nucleons, alphas, and heavy nuclei, neutrino 
absorption on nucleons, neutrino-electron inelastic scattering, 
electron-positron annihilation to neutrino-antineutrino pairs, as 
well as nucleon-nucleon bremsstrahlung.
The default NuLib rates are used (see 
Ref.~\cite{NuLibOnline} for further documentation).
For every EOS, the neutrino opacity tables are created using
Refs.~\cite{OConnor2015,NuLibOnline}.  Because spherically symmetric
simulations do not explode for the chosen progenitor, we artificially
increase the energy deposition by neutrinos in the gain layer by means
of a heating factor. We emphasize that in multi-dimensional
simulations the PNS is spherical even if there is convection inside
(see, e.g., Ref.~\cite{Buras2006}). The use of spherically
symmetric simulations to study the PNS behavior and its sensitivities
to EOS parameters is therefore useful. However, the impact of convection
needs to be also investigated as it may be relevant during the
cooling phase~\cite{Roberts2012}. The heating factor was chosen to
produce an explosion for the LS220$^\dagger$ EOS.

\begin{figure}[t]
\centering
\includegraphics[width=0.95\columnwidth,clip=]{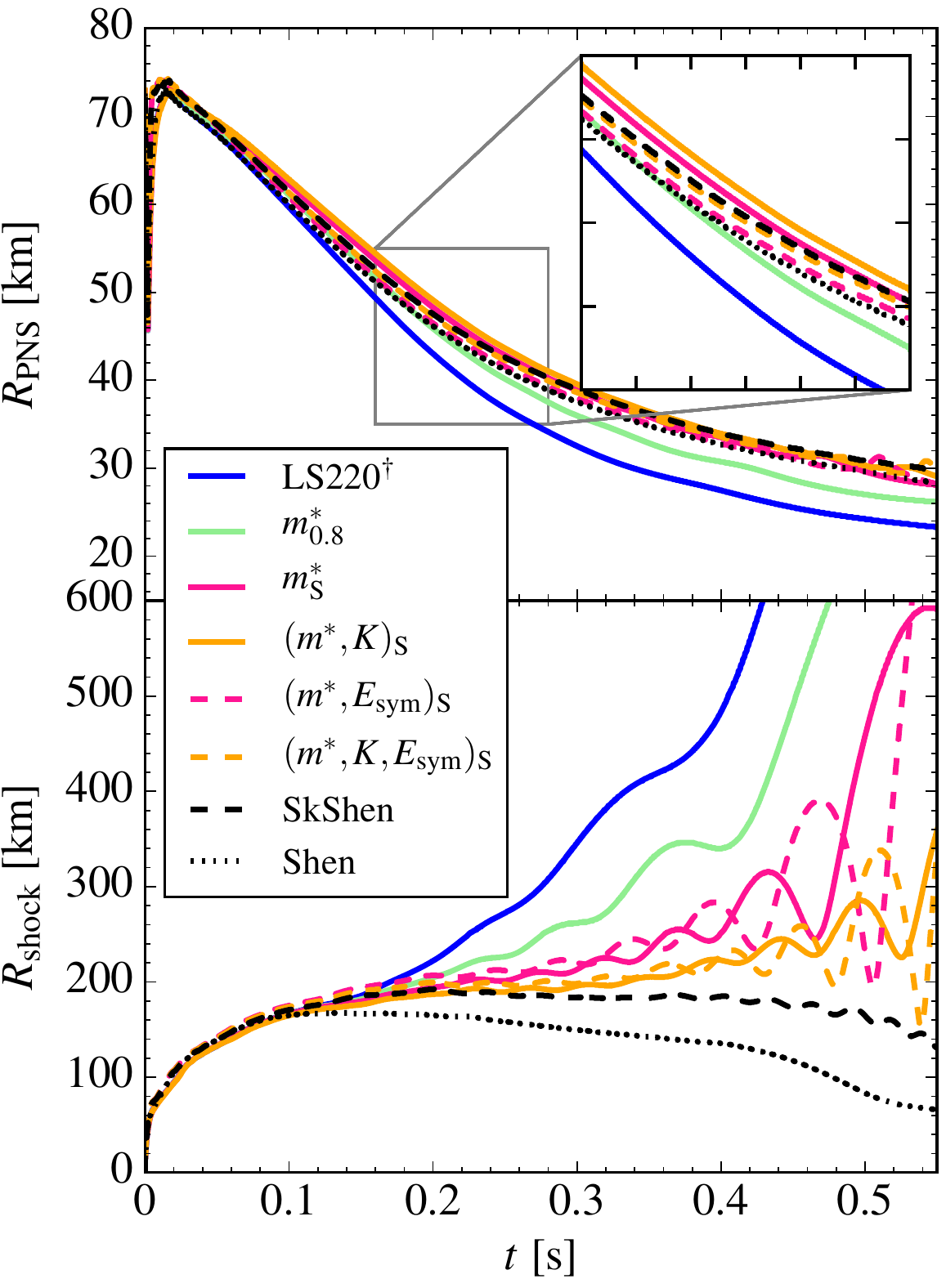}
\caption{Evolution of PNS radius (upper) and shock radius (lower panel) for
supernova simulations of a 15~$M_\odot$ progenitor based on EOSs with 
different microphysics properties ranging from LS220$^\dagger$ to Shen (as in the legend, 
for details see text). The shock presents an oscillatory
behaviour  for few models that are weakly exploding. This is due to
the competition between accretion and explosion and it is
overestimated in one-dimensional simulations.\label{fig:pns}}
\end{figure}

\begin{figure}[t]
\centering
\includegraphics[width=0.95\columnwidth,clip=]{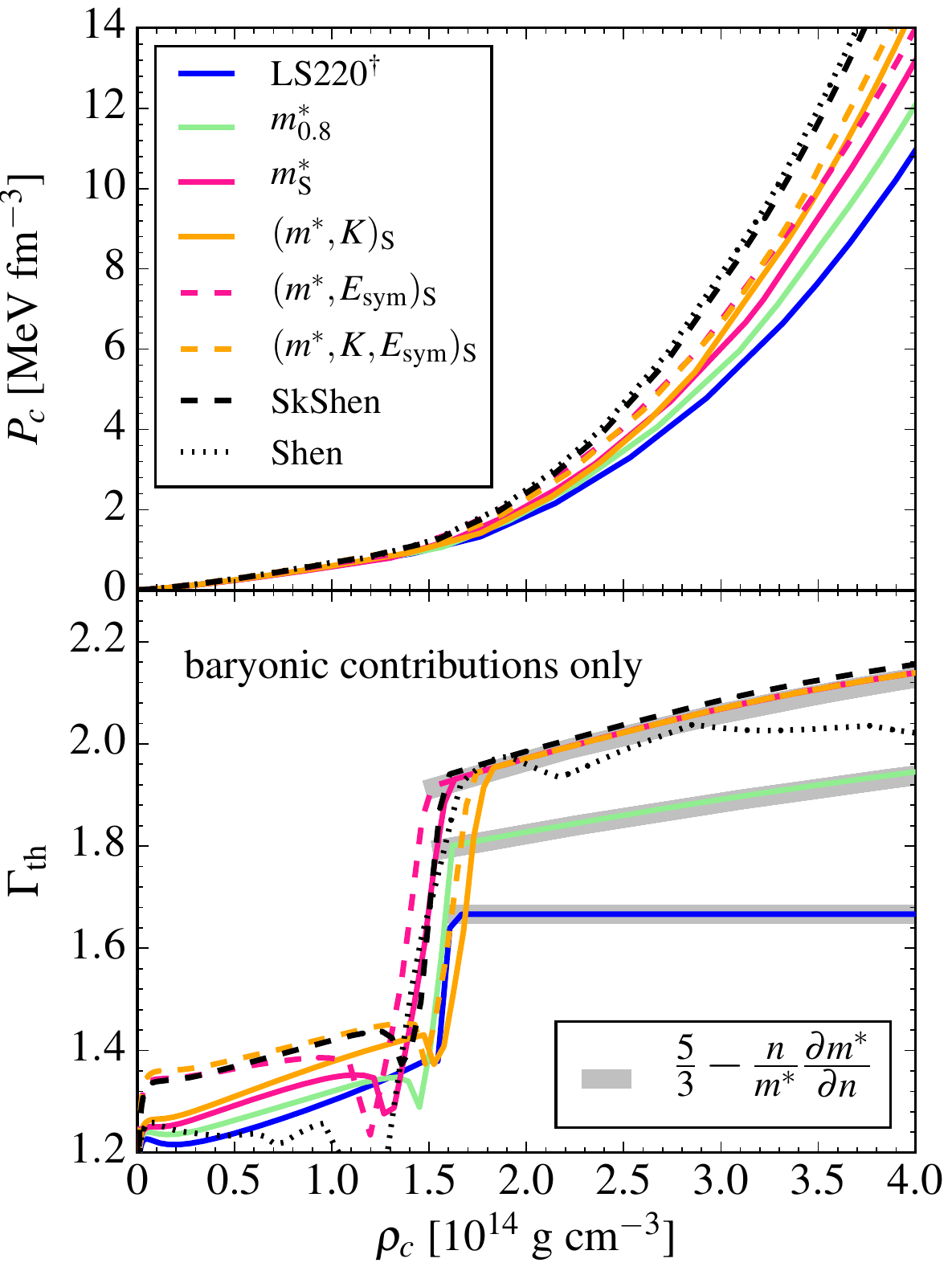}
\caption{Central pressure (upper) and thermal index $\Gamma_\mathrm{th}$
(lower panel) as function of central density for the same simulations and EOSs as in Fig.~\ref{fig:pns}.  The results for $\Gamma_\mathrm{th}$ are
given for the baryonic contributions only, and are compared against
$\Gamma_\mathrm{th}$ of a noninteracting gas
of nonrelativistic fermions with density-dependent $m^*$, Eq.~(\ref{eq:Gth_theo}), shown as
thick gray bands at high densities for $m^* = m$, $m^*_{0.8}$, and
$m^*_\mathrm{S}$.\label{fig:Pc_GammaTh}}
\end{figure}

{\it Proto-neutron star and shock behavior.--} Figure~\ref{fig:pns}
shows the evolution of the PNS radius (defined as the position where
the density is $10^{11} \gcmiq$) and shock radius post bounce (at
$t=0$~s) based on the constructed EOSs that change the microphysics
systematically from LS220 to Shen.  The upper panel of
Fig.~\ref{fig:pns} shows the critical impact of the
effective mass on the PNS behavior, where $m^*$ mainly determines
whether the contraction is faster (LS220$^\dagger$ with $m^*=m$), intermediate
($m^*_{0.8}$), or slower (all other EOSs with $m^*_\mathrm{S}$). As
discussed in more detail later (see Fig.~\ref{fig:Pc_GammaTh}), when
the effective mass is reduced, the pressure increases ($P\sim
1/m^*$), and the observed behavior can be clearly traced to the
thermal effects that depend on the effective mass and its
density dependence. As shown by the shock evolution in the lower panel
of Fig.~\ref{fig:pns}, this has a direct impact on the explosion. When
the effective mass is larger (LS220$^\dagger$) and the PNS contraction faster,
the neutrino energies are higher. This leads to an earlier explosion
and larger shock radii right after the explosion sets
in.\footnote{Note that the shock follows the evolution of the 
PNS~\cite{Janka2012review} and thus the shock radius is smaller 
when the PNS contraction is stronger. This is found in non-exploding 
models. However, if the neutrino energies become high enough, then
the explosion and expansion of the shock are stronger than this trend
to follow the PNS evolution.}

\begin{figure}[t]
\centering
\includegraphics[width=0.95\columnwidth,clip=]{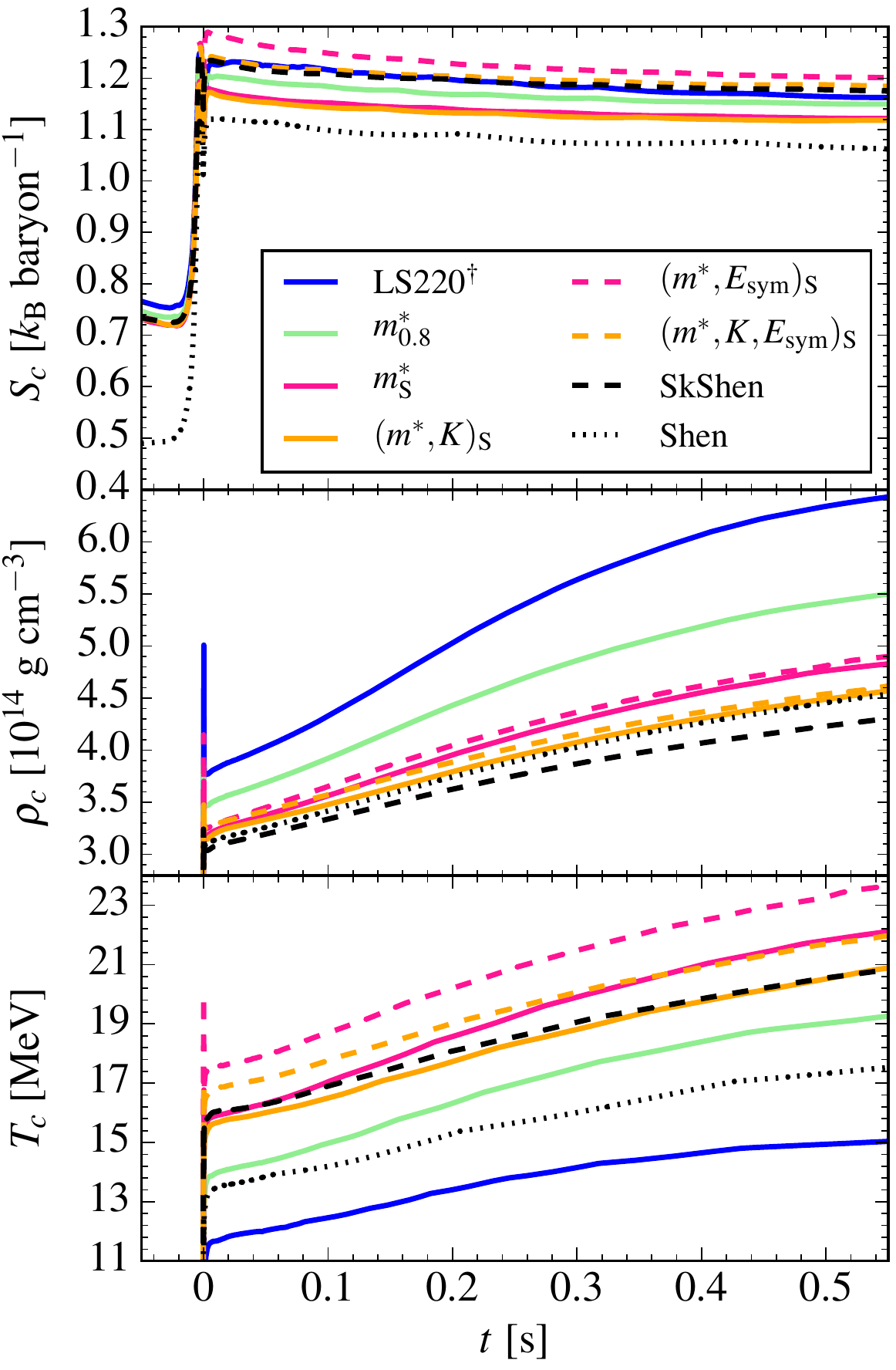}
\caption{Evolution of central values for the entropy (upper), density
(middle), and temperature (lower panel) for the same simulations and
EOSs as in Fig.~\ref{fig:pns}.\label{fig:entropy_temperature}}
\end{figure}

The impact of the incompressibility can be analyzed by comparing the
EOS with $m^*_\mathrm{S}$ and $(m^*,K ) _\mathrm{S}$ in
Fig.~\ref{fig:pns} (see also Refs.~\cite{Couch2013, Suwa2013}). The
larger Shen incompressibility implies a higher pressure, which leads
to a slightly larger PNS radius. However,
this impact is much smaller compared to the changes due the effective
mass. The symmetry energy has also a minor impact on the PNS and shock
evolution, as is evident by comparing the EOS with $m^*_\mathrm{S}$ to
$(m^*,E_\mathrm{sym} ) _\mathrm{S}$ and the EOS with $(m^*,K )
_\mathrm{S}$ to $(m^*,K,E_\mathrm{sym} ) _\mathrm{S}$.
The symmetry energy variation of the PNS
evolution is mainly due to the different conditions during collapse
that result in a slightly larger electron fraction post-bounce for the
models with higher symmetry energy (see also later,
Fig.~\ref{fig:entropy_temperature}). The SkShen EOS is as 
similar as possible to the Shen EOS in terms
of the nuclear physics input; however the underlying framework is
still different. The evolution of the shock is affected by several 
aspects besides the PNS contraction: different neutrino energies 
and luminosities because of variations in the interior PNS properties
(see Fig.~\ref{fig:entropy_temperature}), bounce (initial time and position of the shock), as
well as accretion evolution (see Refs.~\cite{Hempel.etal:2012,Couch2013} for a discussion).
All this combined with the chosen heating factor contributes to the 
differences in shock evolution for the SkShen and Shen EOS.
Nevertheless, qualitatively both SkShen and Shen evolutions are now 
more similar, especially for the shock behavior with an unsuccessful 
explosion.

{\it Impact of EOS on PNS interior.--} To further study the impact of
the different EOS parameters, we show the evolution of the central
entropy, density, and temperature before and after bounce in
Fig.~\ref{fig:entropy_temperature} for the various EOSs considered. The
central entropy (upper panel) only slightly depends on the effective
mass. Note that the low central entropy obtained with the Shen EOS is
due to the absence of kinetic entropy of
nuclei~\cite{Hempel.etal:2012}. The symmetry energy determines the
electron fraction and entropy during collapse and after
bounce~\cite{BETHE1979,Mazurek1979}. As shown in the top panel of
Fig.~\ref{fig:entropy_temperature}, the EOSs with the lower symmetry
energy have lower entropy and the post-bounce central electron
fraction is $Y_{e,c} \approx 0.27$ compared to $Y_{e,c} \approx 0.30$
obtained for the higher Shen symmetry energy.

The central density (middle panel of
Fig.~\ref{fig:entropy_temperature}) follows the effective mass
hierarchy, because the pressure scales as $P_c \sim 1 / m^*$; this can
further be seen in the top panel of Fig.~\ref{fig:Pc_GammaTh}. The PNS
radii in Fig.~\ref{fig:pns} approximately follow the same hierarchy as
the central density. Increasing the incompressibility and lowering
the saturation density yields even higher central pressures, which in
turn lowers the central density reached in the simulation.

The central temperature (lower panel in
Fig.~\ref{fig:entropy_temperature}) is affected by changes in the
effective mass as well as the symmetry energy.  This can be understood
considering that the entropy is approximately constant and independent
of the EOS, and assuming a Fermi liquid theory scaling, $S_c \sim m^*
T_c/\rho_c^{2/3}$~\cite{BaymPethick}.  Reducing the effective mass
thus increases the central temperature. Moreover, the larger value
for the symmetry energy in the $(m^*, E_\mathrm{sym})_\mathrm{S}$,
$(m^*, K, E_\mathrm{sym})_\mathrm{S}$, and SkShen EOS increases the
central entropy and thus the central temperature. Similarly, the effect
of the incompressibility can be understood through its impact on the
central density. For the simulation based on the Shen EOS, the
temperature is lower as expected from the entropy behavior discussed above.

{\it Diagnosing thermal effects.--} We have seen that the EOS
impacts the interior of the PNS and thus the PNS
contraction. Because $P_c \sim 1 / m^*$, we find larger central
pressures for smaller $m^*$ as shown in the top panel of
Fig.~\ref{fig:Pc_GammaTh}.  The incompressibility determines the slope
of the pressure, resulting in stiffer EOSs for the larger Shen
incompressibility. In addition, the larger Shen symmetry energy yields
even higher pressures, as this correlates with the $L$ parameter. The
SkShen EOS results in the largest pressures of all our EOSs.
This is due to the smaller saturation
density, which leads to a larger pressure compared to an EOS starting
from a higher $n_0$ (where $P=0$).

At the mean-field level in uniform matter, the thermal nucleonic
contributions to the EOS are completely determined by the effective
mass within the LS Skyrme functionals.  In this approximation, the
thermal index $\Gamma_\mathrm{th}$ of a noninteracting gas of
nonrelativistic fermions with density-dependent $m^*$ is given by
(see, e.g., Ref.~\cite{Cons15thermal})
\begin{equation}
\Gth = \frac{5}{3} - \frac{n}{m^*} \frac{\partial m^*}{\partial n} \,. \label{eq:Gth_theo}
\end{equation}
We calculate $\Gth$ from our simulations for all constructed EOSs by
separating the pressure $P$ and energy density $\varepsilon$ into a
cold and thermal (th) part following Ref.~\cite{Bauswein10thermal},
\begin{equation}
\Gth = 1 + \frac{P_\mathrm{th}}{\varepsilon_\mathrm{th}} = 
1+\dfrac{P-P_\mathrm{cold}}{\varepsilon-\varepsilon_\mathrm{cold}} \,,
\end{equation}
where we extract $P_\mathrm{cold}$ and $\varepsilon_\mathrm{cold}$
from the EOS table at the minimal temperature of $T=0.01\MeV$. This is
shown for the baryonic contributions only in the lower panel of
Fig.~\ref{fig:Pc_GammaTh}. At high densities, we also compare this
against $\Gamma_\mathrm{th}$ of Eq.~(\ref{eq:Gth_theo}) shown as thick
gray bands for the three different effective mass scenarios. The
agreement is excellent, showing that a decreasing effective mass leads
to a larger $\Gamma_\mathrm{th}$ and thus a larger thermal
contribution to the pressure. Note that SkShen has the same $m^*$
value at $n_0$, but a smaller saturation density, leading to a
slightly larger $\Gamma_\mathrm{th}$ than the other $m^*_\mathrm{S}$
EOSs. The remaining differences to the Shen EOS are attributed to the
underlying relativistic mean-field formalism used. Below the
phase transition, $\rho_c \lesssim 1.7 \gcmiq$, matter is no longer
uniform and also clustering affects the thermal index.

\begin{figure}[t]
\centering
\includegraphics[width=0.85\columnwidth,clip=]{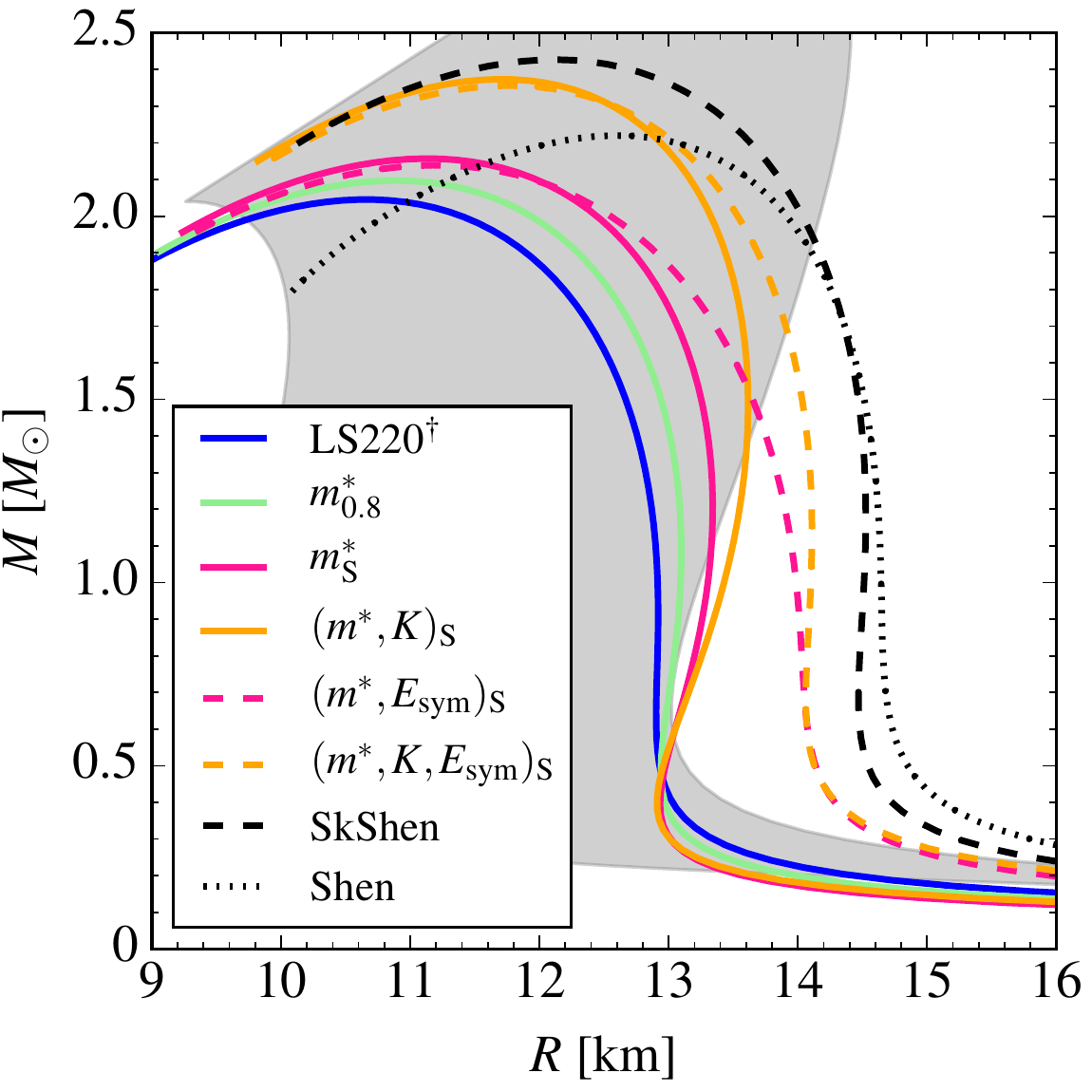}
\caption{Mass-radius relation for cold ($T = 0.1\MeV$) neutron stars in beta
equilibrium for the various EOSs considered in this work. For
comparison, we show the gray band from Ref.~\cite{Hebe13ApJ} based on
chiral EFT calculations up to saturation density
and a general extension to high densities.\label{fig:MR}}
\end{figure}

{\it Cold neutron stars.--} Finally, we calculate the mass-radius
($M$--$R$) relations for cold neutron stars to verify that the
constructed EOSs give reasonable modifications to the $M$--$R$
relation. To this end, we solve the Tolman-Oppenheimer-Volkoff
equations~\cite{Oppe39tov} for $T=0.1 \MeV$ and vanishing neutrino
chemical potential. The results are shown in Fig.~\ref{fig:MR}.  All
new EOSs are able to support a two-solar-mass neutron
star~\cite{Anto13PSRM201,Fonseca2016}.  Because the neutron star
radius scales with the pressure of neutron matter at saturation
density~\cite{LatPrakash2007,LattimerLim}, the radius and also the
maximum mass in Fig.~\ref{fig:MR} increase with decreasing $m^*$ and
larger incompressibility due to the larger pressures. Moreover,
because the symmetry energy correlates with the $L$ parameter in the
LS Skyrme model, we find that the radius increases significantly once
the EOS used the large Shen symmetry energy. As the
$L$ parameters for the EOSs constructed here are high compared to
chiral EFT calculations (see Table~\ref{tab:parameters}), the EOSs
considered lie towards larger radii compared to the gray band from
Ref.~\cite{Hebe13ApJ} (see Fig.~\ref{fig:MR}) based on chiral EFT
calculations combined with a general extension to high
densities. Moreover, it is reassuring that the SkShen EOS is
similar to the relativistic energy-density functional based Shen EOS,
once the same EOS parameters are used. This shows that indeed
the physical properties are the important microphysics input and not
the detailed scheme of the functional.

In summary, we have investigated core-collapse supernova simulations
based on a range of EOSs by varying the nucleon effective mass,
incompressibility, symmetry energy, and nuclear saturation point
systematically from LS220 to Shen. All constructed EOS tables are available online \cite{online_tables}. In
particular, we have shown that the effective mass has a decisive
effect on the PNS contraction, with larger effective masses leading to
a smaller thermal contribution to the pressure and thus a more rapid
contraction. This aids the shock evolution to a faster explosion. By
varying the EOS from LS220 to Shen, we were able to systematically
step between these two commonly used EOSs and with SkShen show why
the Shen EOS does not result in a successful explosion. While LS220$^\dagger$ was the EOS with the largest effective mass considered in this work, ab initio calculations of the EOS suggest that the effective mass can even increase to $m^* > m$ at higher densities due to contributions
from correlations and three-nucleon forces~\cite{Carb19Gammath}.  The
effects also increased the radius of a cold $1.4~M_\odot$ neutron star
from 12.8~km for LS220$^\dagger$ to 14.6~km for Shen, leading to a larger
maximum mass as well. However, the EOS variation observed for the hot
PNS radius clearly follows the behavior of the thermal effects
diagnosed through the thermal index. Future work will include the
construction of a range of EOSs based on existing and new chiral EFT
constraints as well as further astrophysics explorations including
also multi-dimensional simulations.
	
\begin{acknowledgments}

We thank A. Carbone, S. Greif, K. Hebeler, C. Mattes, E. O'Connor, and S. Couch
for useful discussions.  This work was supported by the Deutsche Forschungsgemeinschaft (DFG, German Research Foundation) -- Project-ID 279384907 -- SFB 1245 and
the European Research Council Grant No.~677912 EUROPIUM.
		
\end{acknowledgments}
	
\bibliographystyle{apsrev4-1}
\bibliography{bibliography}
	
\end{document}